\documentclass[final,3p,times]{elsarticle}

\usepackage{epsfig}
\usepackage{graphicx}
\usepackage{epstopdf}
\usepackage{color}
\usepackage{amssymb}
\usepackage{amsthm}

\journal{Annals of Physics}

\begin{document}

\begin{frontmatter}



\title{Berry-phase-based quantum gates assisted by transitionless quantum driving}
\author{Shifan Qi}
\author{Jun Jing\corref{cor1}}
\ead{jingjun@zju.edu.cn}
\cortext[cor1]{Corresponding author}
\address{Zhejiang Province Key Laboratory of Quantum Technology and Device, Department of Physics, Zhejiang University, Hangzhou 310027, Zhejiang, China}

\begin{abstract}
We propose a novel proposal for geometric quantum gates using three- or two-level systems, in which a controllable variable, the detuning between the driving frequency and the atomic energy spacing, is introduced to realize geometric transformations. In particular, we can have two instantaneous eigenstates with opposite eigenvalues constituting a closed loop in the parameter space. The accumulated dynamical phase is then exactly cancelled when the loop is completed, which is beyond the traditional parallel-transport restriction. We apply the transitionless quantum driving, which renders revisions in both amplitudes and phases of the driving fields, to enhance the speed and the fidelity of geometric transformation in both universal single-qubit gates and nontrivial double-qubit gates. Gate fidelity under decoherence is also estimated.
\end{abstract}



\end{frontmatter}


\section{Introduction}

Quantum computation is believed to be more superior than its classical counterpart in solving certain problems, such as factoring large integers~\cite{inte} and searching unsorted databases~\cite{data}. In order to implement quantum computation, it is prerequisite to have a universal set of single-qubit gates and a nontrivial double-qubit gate with a high fidelity. Nevertheless, it is always hard to realize reliable and robust quantum gates, because any quantum system or platform is inevitably subjected to the errors arising from inaccurate control and external environments. The quantum gates based on the geometrical phase~\cite{berry} are robust against control errors, since the geometric phase is exclusively determined by the global properties of the transformation paths and independent of the transformation details~\cite{ph1,ph2,ph3,ph4}.

Many schemes of adiabatic or nonadiabatic geometric gates using Berry or non-Abelian~\cite{na1} phase have been proposed~\cite{ge1,ge2,ho1,ho2,ho3,ho4,ho5,tg1,tg2,tg3,ld,nag,ra3}. Many experimental platforms have also been devoted to the geometric quantum computation, including the Rydberg atoms, the trapped ions~\cite{ti1}, nuclear magnetic resonance~\cite{nm1,nm2}, superconducting circuits~\cite{su1,su2}, and nitrogen-vacancy centers in diamond~\cite{nv1,nv2}. The adiabatic geometric gates require that the quantum system undergoes an adiabatic evolution, which usually takes a long time to complete the desired loop in the parameter space. A long time of evolution is harmful to the robustness of the geometric quantum computation, for it will gradually reduce the execution efficiency and amplify the decoherence. It is obviously negative to achieve high-fidelity quantum gates. In additional, the nonadiabatic Berry phases and non-Abelian holonomies allow for high-speed quantum gate, but these gates are sensitive to the systematic errors in the control Hamiltonian~\cite{ng1,ng2}.

Therefore it is required to shorten the evolution time while retaining the adiabatic passage~\cite{track} or maintaining the quantum system as the desired instantaneous eigenstates at the two ends of the evolution. Plenty of superadiabatic approaches~\cite{sta1,sta2,sta3,sta4} have been proposed to dynamical systems, such as, the transitionless quantum driving (TQD)~\cite{cd1,cd2,cd3,sta1}, shortcuts to adiabaticity using Lewis-Riesenfeld invariant~\cite{lr1,lr2,lr3}, dressed-state-based inverse engineering~\cite{ie} and noise-induced adiabaticity~\cite{ns,co,noiseex}, to name a few. The TQD approach was proposed in the first decade of this century, which is useful for both theoretical proposals and practical applications. Its formulation is to add an ancillary Hamiltonian to the original Hamiltonian. Namely, if the original time-dependent Hamiltonian $H(t)$ is formally expressed in the spectral representation as $H(t)=\sum_{n}E_n(t)|n(t)\rangle\langle n(t)|$, then the ancillary term, also named counterdiabatic (CD) term can be written as~\cite{sta1}
\begin{eqnarray}\label{Hcd}
H_{CD}(t)=i\sum_n\left[1-|n(t)\rangle\langle n(t)|\right]|\dot{n}(t)\rangle\langle n(t)|,
\end{eqnarray}
where $E_n(t)$'s and $|n(t)\rangle$'s are respectively the instantaneous eigenvalues and eigenstates, and $|\dot{n}(t)\rangle$'s are the derivatives of the instantaneous eigenstates with respect to time. The approach has been applied in the realization of geometric gates~\cite{stag1,stag2,sta5,sta6,sta7}. And they are believed to be not only remarkably fast in running period but also highly robust against control parameter variations~\cite{sta1,cd2,cd3}.

In this work, we propose a novel scheme of superadiabatic geometric gates assisted by the TQD, which provides a promising way to realize high-fidelity geometric quantum computation. In some of the traditional schemes~\cite{nag,stag2,stag1}, the time-independent off-resonant driven three-level systems are employed for realizing the single-qubit gates. In contrast, our scheme applies a controllable time-dependent detuning between the driving frequency and the atomic level-splitting as an additional variable to cancel the dynamical phase accumulated during the whole closed path in the parameter space. The payback of the overhead is that our scheme can go out of the parallel-transport restriction, then is not susceptible to the systematic errors. Comparing to the superadiabatic scheme~\cite{stag2} in which special gates, e.g., the $\sigma_y$ gate, has to be realized by double loops, we introduce the phase difference between the complex Rabi frequencies of the two driving lasers to implement a universal single-qubit gate by a single loop. It is also the reason that we do not employ a more simple model of off-resonant-driven two-level systems~\cite{stag1}. At the same time, we have perfected the conditions that must be met to eliminate the dynamic phase. In addition, we have promoted the approach to the preparation of the nontrivial double-qubit gate. Our scheme takes the advantage of both the speed in running time and the robustness against control parameter variations~\cite{cd2,cd3} over the traditional schemes for the double-qubit gate~\cite{ra3,tg2,tg3}.

The rest of the work is so arranged as follows. Section~\ref{one} is devoted to establish a universal set of single-qubit gates based on a three-level system. In our proposal, the type of a quantum gate is determined by the phase difference between the complex Rabi frequencies of the two driving lasers. TQD is then implemented to improve the adiabaticity of the geometric transformation as well as the gate efficiency and fidelity. We also analysis the average transformation-fidelity over various initial states under decoherence. In Sec.~\ref{two1}, similar discussion is extended to the nontrivial double-qubit gates founded on two two-level systems. In Sec.~\ref{two2}, we propose an alternative scheme for the double-qubit gates, in which a two-level system couples simultaneously to two three-level systems. We conclude our work in Sec.~\ref{conc}.

\section{Universal single-qubit gates}\label{one}

\begin{figure}[htbp]
\centering
\includegraphics[width=0.3\textwidth]{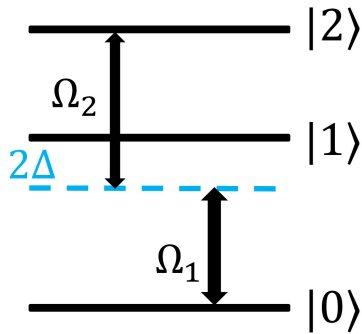}
\caption{(Color online) Diagram for a three-level system under driving. The stable ground state $|0\rangle$ is coupled to the intermediate state $|1\rangle$ by an off-resonant laser with Rabi frequency $\Omega_1$. The intermediate state $|1\rangle$ and the excited state $|2\rangle$ is driven by another off-resonant laser with Rabi frequency $\Omega_2$. $2\Delta$ is the detuning between the driving frequency and the energy spacing between states $|0\rangle$ and $|1\rangle$. }\label{singlegate}
\end{figure}

Consider a three-level system consisted of a stable ground state $|0\rangle$, an intermediate state $|1\rangle$, and a highly excited state $|2\rangle$. $|0\rangle$ is coupled to $|1\rangle$ by an off-resonant laser with Rabi frequency $\Omega_1$. States $|1\rangle$ and $|2\rangle$ are driven by another off-resonant laser with Rabi frequency $\Omega_2$ (see the diagram in Fig.~\ref{singlegate}). In the unit of $\hbar\equiv1$, the system Hamiltonian can be expressed by~\cite{H}
\begin{eqnarray}\label{H0one}
H_0(t)=\omega_1|1\rangle\langle1|+\omega_2|2\rangle\langle2|+\left[\Omega_1(t)e^{i\Xi_1(t)}|0\rangle\langle1|+\Omega_2(t)e^{-i\Xi_2(t)}|2\rangle\langle1|+h.c.\right].
\end{eqnarray}
Here the energy of level $|0\rangle$ is assumed as $\omega_0=0$ with no loss of generality and $\Xi_n(t)\equiv\int_0^tds\xi_n(s)$, $n=1,2$. $\omega_n$ is the energy of level $|n\rangle$. $\Omega_n(t)$ and $\xi_n(t)$ are respectively the complex time-dependent Rabi frequency and the real time-dependent driving frequency corresponding to the driving term between $|n-1\rangle$ and $|n\rangle$. $\xi_1(t)$ and $\xi_2(t)$ satisfy the condition that $\xi_1(t)+\xi_2(t)=\omega_2$.

Turn to the rotating frame with respect to $U_0(t)=\exp\{i\Xi_1(t)|1\rangle\langle1|+i[\Xi_1(t)+\Xi_2(t)]|2\rangle\langle2|\}$, the Hamiltonian~(\ref{H0one}) can be rewritten as
\begin{eqnarray}\label{Hone}\nonumber
&H(t)=U_0(t)H_0(t)U_0^\dagger(t)-iU_0(t)\dot{U}_0^\dagger(t) \\
&=\left[\Omega_1(t)|0\rangle\langle1|+\Omega_2(t)|2\rangle\langle1|+h.c.\right]-2\Delta(t)|1\rangle\langle1|,
\end{eqnarray}
where $2\Delta(t)\equiv\xi_1(t)-\omega_1$. The two complex Rabi frequencies can be parameterized by
\begin{eqnarray}\nonumber
&\Omega_1(t)=\Omega(t)\sin(\theta/2)e^{-i\phi}, \\  \label{omega}
&\Omega_2(t)=\Omega(t)\cos(\theta/2)e^{-i\phi+i\psi},
\end{eqnarray}
with $\Omega(t)$ being real and two controllable phase parameters $\phi$ and $\psi$. Thus the Hamiltonian becomes
\begin{eqnarray}
\label{Hone2}
H(t)=\Omega(t)\big[\sin(\theta/2)e^{-i\phi}|0\rangle\langle1|+\cos(\theta/2)e^{-i\phi+i\psi}|2\rangle\langle1|+h.c.\big]-2\Delta(t)|1\rangle\langle1|.
\end{eqnarray}
In a frame spanned by a new orthonormal set $\{|b\rangle, |1\rangle, |d\rangle\}$, where the bright and dark states are respectively defined by
\begin{eqnarray}\nonumber
|b\rangle=\sin(\theta/2)|0\rangle+\cos(\theta/2)e^{i\psi}|2\rangle, \\ \label{bd}
|d\rangle=\cos(\theta/2)|0\rangle-\sin(\theta/2)e^{i\psi}|2\rangle,
\end{eqnarray}
one can express the Hamiltonian~(\ref{Hone2}) as
\begin{equation}\label{Hone3}
H(t)=\Omega(t)\left[e^{-i\phi}|b\rangle\langle1|+h.c.\right]-2\Delta(t)|1\rangle\langle1|.
\end{equation}
It is clear that $|d\rangle$ is decoupled from $|b\rangle$ and $|1\rangle$, which means that $|d\rangle$ remains unchanged during the system transformation. So the Hamiltonian~(\ref{Hone3}) can be effectively expressed in the standard two-state space spanned by $\{|b\rangle,|1\rangle\}$. Up to an identity operator with a factor $\Delta(t)$, we have
\begin{eqnarray}\label{Hb}
H(t)=\left[\begin{array}{cc}
\Delta(t) & \Omega (t)e^{-i\phi}\\
\Omega(t)e^{i\phi} & -\Delta(t)
\end{array}\right].
\end{eqnarray}
$\Omega(t)$ and $\Delta(t)$ can be further parameterized by $\Omega(t)=E(t)\sin\varphi(t)$ and $\Delta(t)=E(t)\cos\varphi(t)$, respectively. Then the instantaneous eigenstates and eigenvalues of the Hamiltonian~(\ref{Hb}) can be written as
\begin{eqnarray}\label{eigenstate}\nonumber
|E_+(t)\rangle=\cos[\varphi(t)/2]|b\rangle+\sin[\varphi(t)/2]e^{i\phi}|1\rangle, \\
|E_-(t)\rangle=-\sin[\varphi(t)/2]e^{-i\phi}|b\rangle+\cos[\varphi(t)/2]|1\rangle,
\end{eqnarray}
and $E_{\pm}(t)=\pm E(t)$ respectively.

The quantum system is supposed to evolve from $t=0$ to $t=T$. Our geometric process is actually designed as a concatenated circle consisted of two piece-wisely adiabatic passages, $0\rightarrow t_f$ and $t_f\rightarrow T$ with $t_f=T/2$, to achieve the evolution from $|b\rangle$ at $t=0$ to $e^{i(\beta+\eta)}|b\rangle$ at $t=T$, where $\beta$ and $\eta$ are respectively the accumulated dynamical and geometric phases.

In particular, the system can evolve as $|E_+(t)\rangle$ during the first-half period $t\in[0,t_f)$ and is then relayed by $|E_-(t)\rangle$ during the second-half part $t\in[t_f,T]$. The parameter $\phi$ from Eq.~(\ref{omega}) is set as $\phi_1$ during the first-half part and $\phi_2$ during the second-half part of evolution, where $\phi_1$ and $\phi_2$ are controllable constants. Due to the fact that $E_+(t)=-E_-(t)=E(t)$, and the assumption that $E(t)$ satisfies
\begin{equation}\label{EC}
E(t)=E(t+t_f)\quad {\rm or} \quad E(t)=E(T-t),
\end{equation}
then the accumulated dynamical phase exactly vanishes when completing the whole loop,
\begin{equation}\label{phi1phi2}
\beta=-\left[\int_0^{t_f}dtE_+(t)+\int_{t_f}^TdtE_-(t)\right]=0,
\end{equation}
which can then be safely ignored in the following discussions. With respect to the whole geometric transformation, to allow the evolution process determined by the adiabatic theorem
\begin{eqnarray}\nonumber
|E_+(0)\rangle=|b\rangle \to |E_+(t_f-0^+)\rangle=e^{i\phi_1}|1\rangle, \\ \label{evolution}
e^{i\phi_1}|1\rangle=e^{i\phi_1}|E_-(t_f)\rangle\to e^{i\phi_1}|E_-(T)\rangle=e^{i\eta}|b\rangle,
\end{eqnarray}
the time-dependent parameter $\varphi(t)$ should be conditioned by $\varphi(0)=0$, $\varphi(t_f-0^+)=\pi$, $\varphi(t_f)=0$, and $\varphi(T)=\pi$, where $\eta=\pi+\phi_1-\phi_2$. Note one can design a similar two-stage transformation as above by swapping $|E_+(t)\rangle$ and $|E_-(t)\rangle$ with a proper function $\varphi(t)$.

Therefore throughout the whole period $[0,T]$, the geometric phase is found to be $\eta=\pi+\phi_1-\phi_2$ and the bright state $|b\rangle$ turns to be $e^{i\eta}|b\rangle$. It is known that the dark state remains invariant with time. Thus one can obtain a unitary transformation operator in the subspace spanned by $\{|d\rangle, |b\rangle\}$:
\begin{eqnarray}\label{U}
U=\left[\begin{array}{cc}
1 & 0\\
0 & e^{i\eta}
\end{array}\right]\simeq e^{-\frac{i\eta}{2}(|d\rangle\langle d|-|b\rangle\langle b|)}.
\end{eqnarray}
In the computational subspace spanned by $\{|0\rangle, |2\rangle\}$, we will obtain a universal gate operation for the one-qubit system:
\begin{eqnarray}\label{oneq}\nonumber
U&=&\left[\begin{array}{cc}
\cos\frac{\eta}{2}-i\sin\frac{\eta}{2}\cos\theta & i\sin \frac{\eta}{2}\sin\theta e^{-i\psi}\\
i\sin\frac{\eta}{2}\sin\theta e^{i\psi} & \cos\frac{\eta}{2} +i\sin \frac{\eta}{2}\cos\theta
\end{array}\right]\\
&=&\exp\left(i\frac{\eta}{2}\vec{n}\cdot\vec{\sigma}\right),
\end{eqnarray}
where $\vec{\sigma}$ is a vector of Pauli matrices. On the Bloch sphere, this operation can rotate an {\it arbitrary} unit vector around the axis $\vec{n}=(\sin\theta\cos\psi, \sin\theta\sin\psi, -\cos\theta)$ by an angle $\eta$ in the clockwise direction, where $\theta$ and $\psi$ are parameterized by the two Rabi frequencies in Eq.~(\ref{omega}) and $\eta$ is the geometric phase determined by $\phi_1-\phi_2$ as in Eq.~(\ref{evolution}). A combination of $\theta$ (measuring the relative strengthes of the two driving fields), $\psi$ (the phase difference between the two driving fields) and $\phi_1-\phi_2$ (a phase acting as a controllable constant) corresponds to a specific single-qubit gate. For example, one can set $\theta=\pi$, and $\phi_1=\phi_2$ to obtain the $\sigma_z$ gate up to a global phase.

To ensure the transitionless evolution during the adiabatic passage while shortening the period of the loop presented in Eq.~(\ref{evolution}), we use the TQD by adding an ancillary term $H_{CD}(t)$ to the original Hamiltonian~(\ref{Hone}). According to Eq.~(\ref{Hcd}), the ancillary term can be expressed by
\begin{eqnarray}\label{Hcd}
H_{CD}(t)= \left[\begin{array}{cc}
0 & -i\Lambda(t) e^{-i\phi}\\
i\Lambda(t)e^{i\phi} & 0
\end{array}\right],
\end{eqnarray}
where $\Lambda(t)=[\dot{\Omega}(t)\Delta(t)-\Omega(t)\dot{\Delta}(t)]/[2E^2(t)]$. Then we obtain a modified Hamiltonian $H_{TQD}(t)=H(t)+H_{CD}(t)$, which guarantees the quantum system exactly moving along certain eigenstates presented in Eq.~(\ref{eigenstate}) [Note now the eigenvalues $E_\pm(t)$ are modified accordingly but they are irrelevant to both dynamical and geometric phases]. In particular, now the amplitudes and phases of the two driving fields become
\begin{eqnarray}\label{omegaM}\nonumber
\Omega_1(t)\to\Omega^{TQD}_1(t)=[\Omega(t)-i\Lambda(t)]\sin(\theta/2) e^{-i\phi}, \\
\Omega_2(t)\to\Omega^{TQD}_2(t)=[\Omega(t)-i\Lambda(t)]\cos(\theta/2) e^{-i\phi+i\psi}
\end{eqnarray}
respectively. Then one can repeat the parametric routine from Eq.~(\ref{omega}) to Eq.~(\ref{oneq}) to establish a set of more efficient (we will show this shortly) quantum gates by using the modified Rabi frequencies in Eq.~(\ref{omegaM}).

To measure the performance of the single-qubit gates, we introduce a concept of average effective fidelity $F$, which is defined by~\cite{tg2}
\begin{equation}\label{fone}
F(t)=\frac{1}{4\pi^2}\int^{2\pi}_0\int^{2\pi}_0d\alpha_1d\alpha_2\left|\langle\Psi_U|U(t)|\Psi(0)\rangle\right|^2.
\end{equation}
Here ``average'' means the sampling from the computational subspace, i.e., the initial states of the system are expressed by $|\Psi(0)\rangle=\cos\alpha_1|0\rangle+\sin\alpha_1e^{i\alpha_2}|2\rangle$ with $\{\alpha_1, \alpha_2\}\in[0,2\pi]$. $U(t)$ is the evolution operator directly obtained from the Hamiltonian~(\ref{Hb}) with the designed Rabi frequencies $\Omega_1(t)$ and $\Omega_2(t)$ in Eq.~(\ref{omega}) or the effective Rabi frequencies $\Omega^{TQD}_1(t)$ and $\Omega^{TQD}_2(t)$ in Eq.~(\ref{omegaM}). $|\Psi_U\rangle\equiv U|\Psi(0)\rangle$ is the target state after performing the single-qubit gate $U$ in Eq.~(\ref{oneq}) on $|\Psi(0)\rangle$. In this work, we set~\cite{stag2} $E(t)=1$ and
\begin{equation}\label{varphi}
\varphi(t)=\left\{\begin{array}{ll}
\frac{3\pi t^2}{t^2_f}-\frac{2\pi t^3}{t^3_f}\quad (0\le t <t_f) \\
\frac{3\pi(t-t_f)^2}{t^2_f}-\frac{2\pi(t-t_f)^3}{t^3_f}\quad (t_f \le t\le 2t_f=T)\end{array} \right.
\end{equation}

In the following we will check the effect of our scheme with or without TQD approach for four types of single-qubit gates by calculating the final average fidelity of these gates versus the running time $T$ and the average fidelity dynamics versus time $t$ until a fixed running time $T$. According to Eq.~(\ref{oneq}), when $\theta=\pi$ and $\phi_1=\phi_2$, we have the $\sigma_z$ gate; when $\theta=\pi/2$, $\phi_1=\phi_2$ and $\psi=0$, we have the $\sigma_x$ gate; when $\theta=\pi/2$, $\phi_1=\phi_2$, and $\psi=\pi/2$, we have the $\sigma_y$ gate; and when $\theta=0$ and $\phi_1-\phi_2=-3\pi/4$, we have the $\pi/8$-phase gate or called $T$-gate.

\begin{figure}[htbp]
\centering
\includegraphics[width=0.5\textwidth]{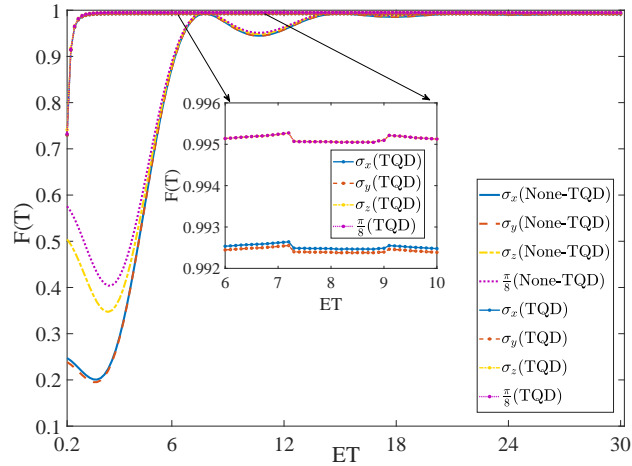}
\caption{(Color online) The effect of the geometric transformation time $T$ on the final average fidelity $F(T)$ of the single-qubit gates. The blue solid lines with or without the circle signs represent the $\sigma_x$ gate when there is a counterdiabatic term or not in the Hamiltonian, respectively. Similarly, the red dashed lines, the orange dashed-dotted lines and the purple dotted lines respectively represent the results of the $\sigma_y$, $\sigma_z$, and $\pi/8$-phase gates.}\label{fidelity1}
\end{figure}

In Fig.~\ref{fidelity1}, we plot the final fidelity $F(T)$ of the four specific single-qubit gates with or without the TQD approach. Under the original Hamiltonian~(\ref{Hb}), the fidelities of all the four gates ($\sigma_x$, $\sigma_y$, $\sigma_z$, $\pi/8$-phase) follow quite similar behaviors. They first lower to some extent and then bounce back until to almost unity at the same cyclic time $T$. After about $ET=7$, they will experience a decline with a much reduced amplitude. Then they are enhanced by a longer $T$ again until maintained at almost unity after $ET\approx17$. It is found that the final average fidelity for all the gates will attain $F\geq0.99$ when $ET\geq15$. While with the help of the ancillary Hamiltonian~(\ref{Hcd}), high fidelity $F\geq0.99$ can be met in a much reduced transformation time about $ET\approx2$. In the inset of Fig.~\ref{fidelity1}, the final average fidelity of the $\sigma_z$ gate (orange dashed-dotted line with circle sign) and the $\pi/8$ gate (purple dotted line with circle sign) follow almost the same dynamics. They are a little bit higher than the fidelities of the other two gates. All of them are greater than $0.99$ after $ET\approx2$.
\begin{figure}[htbp]
\centering
\includegraphics[width=0.5\textwidth]{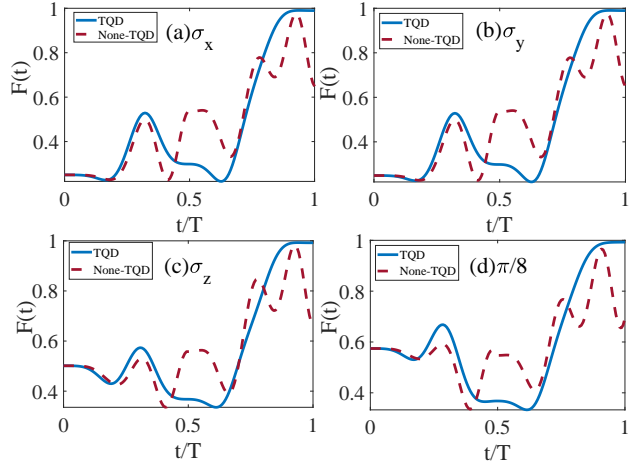}
\caption{(Color online) The dynamics of the average fidelity with a fixed running time $ET=6$ with (blue solid line) and without TQD (red dashed line). (a) for the $\sigma_z$ gate, (b) for the $\sigma_y$ gate, (c) for the $\sigma_x$ gate, (d) for the $\pi/8$-phase gate.}\label{fidelity2}
\end{figure}

The effect of the TQD approach can also be observed from the microscopic evolution of the quantum system. In Fig.~\ref{fidelity2}, we plot the average-fidelity dynamics of the single-qubit gates under a fixed geometric transformation period $ET=6$. In the time domain, the behaviors of the fidelity dynamics are found to be nearly irrespective to the type of the quantum gate. However, the dynamics upon application of the TQD method is quite significant. It will enhance the final fidelity from about $0.85$ to unity in a stable way.

In the open-quantum-system scenario, the fidelity of the geometric transformation is subject to the external decoherence resources. Here we take account both dephasing and dissipation processes into consideration. In the weak-coupling regime, we apply the following Lindblad master equation~\cite{ld}:
\begin{eqnarray}\label{ldone}
\frac{\partial\rho}{\partial t}=-i[H_{TQD}(t),\rho]+\frac{1}{2}\sum_{j \in \{0,2\}}\left[\Gamma_j^-L\left(\sigma^-_{j}\right)+\Gamma_j^zL\left(\sigma^z_{j}\right)\right],
\end{eqnarray}
where $\rho$ is the density matrix of the interested system, $H_{TQD}(t)=H(t)+H_{CD}(t)$ is the TQD-modified Hamiltonian [combining Eqs.~(\ref{Hb}) and (\ref{Hcd})] and $L(A)\equiv2A\rho A^{\dag}-A^{\dag}A\rho-\rho A^{\dag}A$ is the Lindbladian superoperation with the system operator $A$. Here $\sigma_{0}^-\equiv|0\rangle\langle1|$, $\sigma_{2}^-\equiv|1\rangle\langle2|$, $ \sigma^z_{0}\equiv|1\rangle\langle1|-|0\rangle\langle0|$ and $\sigma^z_{2}\equiv|2\rangle\langle2|-|1\rangle\langle1|$. For simplicity, we assume the dissipative rates $\Gamma^-_0=\Gamma^-_2=\gamma_-$ and the dephasing rates $\Gamma^z_0=\Gamma^z_2=\gamma_z$. Here the system is supposed to be prepared at $|0\rangle$. Then ideally the system stays at $|0\rangle$ by the operation of the $\sigma_z$ gate or becomes $|2\rangle$ by the operation of the $\sigma_x$ gate. Accordingly in the presence of decoherence, we investigate the fidelity defined by $F(T)=\langle0|\rho(T)|0\rangle$ for the $\sigma_z$ gate or $F(T)=\langle2|\rho(T)|2\rangle$ for the $\sigma_x$ gate, where $\rho(T)$ represents the time-evolved density operator $\rho(t)$ at the final running moment $T$.

\begin{figure}[htbp]
\centering
\includegraphics[width=0.5\textwidth]{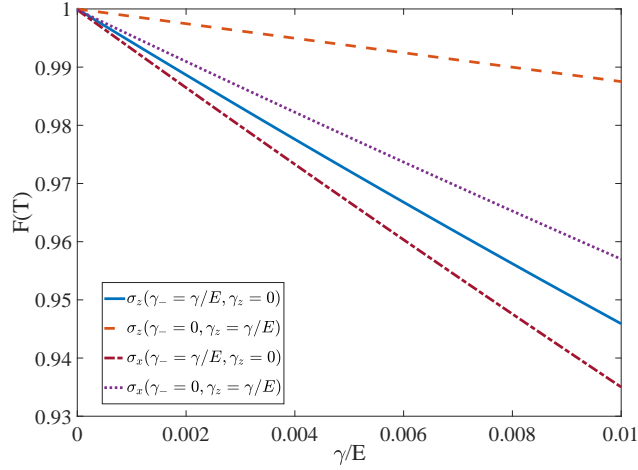}
\caption{(Color online) The decoherence effect on the final fidelity $F(T)$ of system subject to the single-qubit gates. We exam the influence of dissipation indicated by $\gamma_-$ or dephasing indicated by $\gamma_z$ on the $\sigma_z$ and $\sigma_x$ gates. Here the evolution time is fixed as $ET=14$. } \label{singlegamma1}
\end{figure}

\begin{figure}[htbp]
\centering
\includegraphics[width=0.5\textwidth]{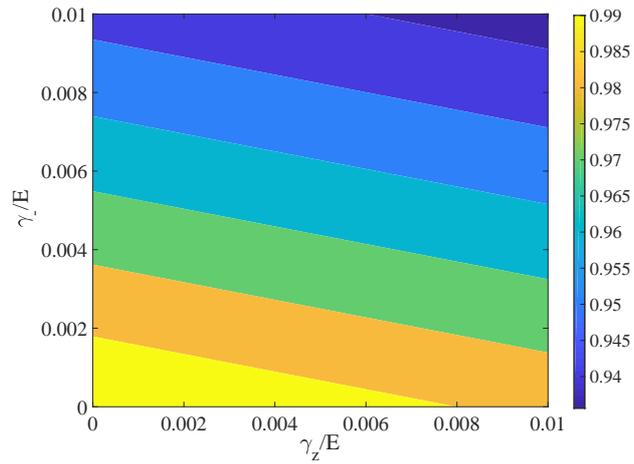}
\caption{(Color online) The decoherence effect on the final fidelity $F(T)$ of system under both dissipation and dephasing. Here we exam the decoherence influence on the $\sigma_z$ gate. The evolution time are fixed as $ET=14$.}\label{singlegamma2}
\end{figure}

In Fig.~\ref{singlegamma1}, we switch on the two decoherence processes in turn for two particular single-qubit gates ($\sigma_z$ and $\sigma_x$). The open system now follows the TQD-assisted geometric transformation while going through a quantum (dissipation or dephasing) channel. We set the cyclic time as $ET=14$. In all the four situations, the fidelity decays linearly with the increasing decay rate. For either gate, the fidelity decay induced by dissipation is quicker than that induced by dephasing under the same amplitude of decay rates $\gamma$. In either decoherence channel, the $\sigma_z$ gate is more robust than the $\sigma_x$ gate. We then demonstrate in Fig.~\ref{singlegamma2} the combined effect from both dissipation and dephasing on the $\sigma_z$ gate. It is shown that the effect on the transformation fidelity from the dephasing noise is about one quarter of that from the dissipation noise in terms of the fidelity decline. The left-bottom triangle area in Fig.~\ref{singlegamma2} indicates the parameter space for maintaining the $\sigma_z$ gate with $F(T)\geq0.99$.

Our scheme to realize a high-fidelity universal single-qubit gate is not restricted to the ladder type, which can be extended to either the $\Lambda$-type or the $V$ type three-level system. It can be implemented in lots of practical systems, such as Rydberg atoms and superconducting circuits. A Rydberg atom is an ordinary atom where one of its electrons, usually the valence electron in an alkali atom, is excited to a state of very high principal quantum number, i.e., a Rydberg state. Neutral atoms in highly excited and long-lived Rydberg states are considered as ideal architectures for quantum computation for they can be used as well-defined two-level or three-level systems. For example, the physical model can be built by $Rb^{87}$ atom with $|0\rangle=|5s_{1/2},F=1,M_F=1\rangle$, $|1\rangle=|5p_{1/2},F=2,M_F=2\rangle$, and $|2\rangle=|58d_{3/2},F=3,M_F=3\rangle$~\cite{Rb}. The relevant Rabi frequencies $\Omega \sim ~1-100 MHz$ and the Rydberg state emission rate $\gamma\sim 10 kHz$~\cite{tg1,Rge}. The superconducting circuit can also be used to implement this scheme. In the transmon qubit, the coherent time is about $\sim 10 \mu s$ and the corresponding dissipation rate is about $\sim 10kHz$, which is one thousand smaller than the Rabi frequencies $\sim 10-100MHz$~\cite{sue}.

\section{Nontrival double-qubit gates I}\label{two1}

\begin{figure}[htbp]
\centering
\includegraphics[width=0.3\textwidth]{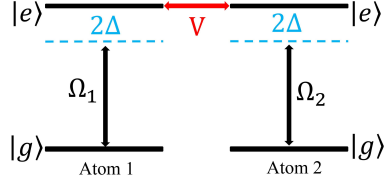}
\caption{(Color online) Diagram for two identical two-level atoms with interaction $V$. For either atom, the stable ground state $|g\rangle$ is coupled to the excited state $|e\rangle$ by an off-resonant laser with Rabi frequency $\Omega_n$, $n=1,2$; and $2\Delta$ is the detuning between the driving frequency and atomic energy splitting. }\label{doublegate1}
\end{figure}

To realize a complete set of gate operations in any circuit model for quantum computation, one needs at least one nontrivial two-qubit gate besides a universal set of one-qubit gates. We now demonstrate how to construct nontrivial two-qubit gates by using two two-level off-resonant atoms with mutual interaction, whose fidelity and efficiency can also be improved by the TQD approach. Consider two identical two-level atoms with energy splitting $\omega$. As shown in Fig.~\ref{doublegate1}, the high-lying excited states of the atoms are coupled through the interaction $V|ee\rangle\langle ee|$. Either atom is under an off-resonant and time-dependent driving $\Omega_n(t)\exp[i\int_0^tds\xi(s)]|g\rangle_n\langle e|+h.c.$, $n=1,2$, with Rabi frequencies $\Omega_n(t)$ parameterized by Eq.~(\ref{omega}) and detuning $2\Delta(t)\equiv\xi(t)-\omega$. Similar to Eq.~(\ref{Hone}), the total system Hamiltonian in the rotating frame with respect to $U_0(t)=\exp[i\int_0^tds\xi(s)(|e\rangle_1\langle e|+|e\rangle_2\langle e|)]$ can be written as~\cite{ra3}
\begin{equation}\label{Htwo}
H(t)=H_1(t)\otimes I_2 +I_1\otimes H_2(t)+V|ee\rangle\langle ee|,
\end{equation}
where the single-atom Hamiltonian describes the interaction between the $n$th atom and the laser, $n=1,2$,
\begin{equation}\label{Hi}
H_n(t)=\Omega_n(t)|g\rangle_n\langle e|+\Omega_n^*(t)|e\rangle_n\langle g|-2\Delta(t)|e\rangle_n\langle e|,
\end{equation}
and $I_n$ denotes the identity operator acting on the $n$th atom. The total Hamiltonian $H(t)$ can then be expressed by
\begin{eqnarray}\label{Hmtwo}\nonumber
H(t)&=&\big[\Omega_1(t)|gg\rangle\langle eg|+\Omega_1(t)|ge\rangle\langle ee|+\Omega_2(t)|gg\rangle\langle ge|+\Omega_2(t)|eg\rangle\langle ee|+h.c.\big]\\
&-&2\Delta(t)|ge\rangle\langle ge|-2\Delta(t)|eg\rangle\langle eg|+\left[V-4\Delta(t)\right]|ee\rangle\langle ee|.
\end{eqnarray}
Then a further unitary transformation with respect to $U_1(t)=\exp(iVt|ee\rangle\langle ee|)$ yields
\begin{eqnarray}\nonumber
\label{Hrwa}
H(t)&=&\big[\Omega_1(t)|gg\rangle\langle eg|+\Omega_1(t)e^{-iVt}|ge\rangle\langle ee|+\Omega_2(t)|gg\rangle\langle ge|+\Omega_2(t)e^{-iVt}|eg\rangle\langle ee|+h.c.\big]\\
&-&2\Delta(t)|ge\rangle\langle ge|-2\Delta(t)|eg\rangle\langle eg|-4\Delta(t)|ee\rangle\langle ee|.
\end{eqnarray}

It is now appropriate to apply a rotating-wave approximation to $H(t)$~(\ref{Hrwa}) when the interaction strength $V$ is much greater than the magnitudes of $\Omega_1(t)$ and $\Omega_2(t)$. At this stage, the fast oscillating terms (with $e^{\pm iVt}$) can be regarded as zero in a moderate time-scale. Then the effective Hamiltonian can be simply expressed in the subspace spanned by $\{|gg\rangle, |ge\rangle, |eg\rangle\}$, which is now decoupled from the rest base $|ee\rangle$,
\begin{equation}\label{Heff}
H_{eff}(t)=\left[\begin{array}{cccc}
0 & \Omega_2(t) & \Omega_1(t)\\
\Omega_2^*(t) & -2\Delta(t) & 0\\
\Omega_1^*(t) & 0 & -2\Delta(t)
\end{array}\right].
\end{equation}
The effective Hamiltonian~(\ref{Heff}) is similar to the Hamiltonian~(\ref{Hone}) for the single-qubit gate. The Rabi frequencies of laser pulses can be still parameterized as those in Eq.~(\ref{omega}). With the same procedure as in section~\ref{one} but in the subspace spanned by $\{|gg\rangle, |b\rangle\equiv\sin(\theta/2)|eg\rangle+\cos(\theta/2)e^{i\psi}|ge\rangle\}$, the effective Hamiltonian reads
\begin{equation}\label{Heffs}
H_{eff}(t)=\left[\begin{array}{cccc}
\Delta(t)& \Omega(t) e^{-i\phi}\\
\Omega(t)e^{i\phi} & -\Delta(t)
\end{array}\right].
\end{equation}
Clearly we have a new dark state $|d\rangle\equiv\cos(\theta/2)|eg\rangle-\sin(\theta/2)e^{i\psi}|ge\rangle$, which is decoupled from both $|gg\rangle$ and $|b\rangle$ and remains invariant with time. Thus the instantaneous eigenstates of Eq.~(\ref{Heffs}) can be immediately obtained from Eq.~(\ref{eigenstate}) by changing the bases:
\begin{eqnarray}\label{eigenstatetwo}\nonumber
|E_+(t)\rangle=\cos[\varphi(t)/2]|b\rangle+\sin[\varphi(t)/2]e^{i\phi}|gg\rangle, \\
|E_-(t)\rangle=-\sin[\varphi(t)/2]e^{-i\phi}|b\rangle+\cos[\varphi(t)/2]|gg\rangle.
\end{eqnarray}
Formally their corresponding eigenvalues are $E_+(t)=E(t)=-E_-(t)$, where $E\equiv\sqrt{\Omega^2(t)+\Delta^2(t)}$ and $\tan\varphi(t)=\Omega(t)/\Delta(t)$, and $E(t)$ is also assumed to satisfy Eq.~(\ref{EC}).

We can still perform a concatenated geometric circle consisted of two piece-wisely adiabatic passages similar to the single-qubit gate in section~\ref{one}. However, there is a dramatic difference between the eigenstates in Eq.~(\ref{eigenstatetwo}) for the current two-qubit gate and those in Eq.~(\ref{eigenstate}) for the single-qubit gate. In Eq.~(\ref{eigenstate}), only one of the two bases constituting either $|E_+(t)\rangle$ or $|E_-(t)\rangle$, i.e., $|b\rangle$, lives in the computational subspace while another one, i.e., $|1\rangle$ does not. In contrast, both bases for the eigenstates in Eq.~(\ref{eigenstatetwo}) live in the computational subspace. So that for the two-qubit gates, both eigenstates would take part in the two stages of adiabatic passages. To have an exactly vanishing dynamical phase while completing the loop as in Eq.~(\ref{phi1phi2}), the two stages are still built up by cutting the whole geometric transformation into half and half, i.e., $t_f=T/2$. Conditioned by $\varphi(0)=0$, $\varphi(t_f-0^+)=\pi$, $\varphi(t_f)=0$, and $\varphi(T)=\pi$ [$\varphi(t)$ can still be chosen as the function in Eq.~(\ref{varphi})], the two stages of the geometric transformation read
\begin{eqnarray}\label{evolutiontwo1}\nonumber
|E_+(0)\rangle=|b\rangle \to |E_+(t_f-0^+)\rangle=e^{i\phi_1}|gg\rangle,  \\
e^{i\phi_1}|gg\rangle=e^{i\phi_1}|E_-(t_f)\rangle \to e^{i\phi_1}|E_-(T)\rangle=e^{i\eta_+}|b\rangle,
\end{eqnarray}
and simultaneously
\begin{eqnarray}
\label{evolutiontwo2}\nonumber
&|E_-(0)\rangle=|gg\rangle \to |E_-(t_f-0^+)\rangle=e^{i(\pi-\phi_1)}|b\rangle, \\
&e^{i(\pi-\phi_1)}|b\rangle=e^{i(\pi-\phi_1)}|E_+(t_f)\rangle \to e^{i(\pi-\phi_1)}|E_+(T)\rangle=e^{i\eta_-}|gg\rangle.
\end{eqnarray}
Here $\eta_{\pm}\equiv\pm(\pi+\phi_1-\phi_2)$, where $\phi_1$ and $\phi_2$ are respectively the constant value of $\phi$ in the first-half and the second-half stage of the period. Importantly, it is sure that the whole cyclic transformation is geometric and the geometric phases accumulated from initial states $|b\rangle$ and $|gg\rangle$ are $\eta_+$ and $\eta_-$, respectively. And obviously the dark state $|d\rangle$ and the double-exciton state $|ee\rangle$ do not participate in the evolution. Thus we can have a transformation operation in the space spanned by $\{|gg\rangle, |d\rangle, |b\rangle, |ee\rangle\}$
\begin{equation} \label{Utwo}
U=\left[\begin{array}{cccc}
e^{i\eta_-} & 0 & 0 & 0\\
0 & 1 & 0 & 0\\
0 & 0 & e^{i\eta_+} & 0\\
0 & 0 & 0 & 1
\end{array}\right].
\end{equation}
In the computational space $\{|gg\rangle, |ge\rangle, |eg\rangle, |ee\rangle\}$, it reads
\begin{equation}\label{twoq}
U=\left[\begin{array}{cccc}
e^{i\eta_-} & 0 & 0 & 0\\
0 & \sin^2\left(\frac{\theta}{2}\right)+\cos^2\left(\frac{\theta}{2}\right)e^{i\eta_+} & \frac{\sin\theta}{2}\left(e^{i\eta_+}-1\right)e^{i\psi} & 0\\
0 & \frac{\sin\theta}{2}\left(e^{i\eta_+}-1\right)e^{-i\psi} & \cos^2\left(\frac{\theta}{2}\right)+\sin^2\left(\frac{\theta}{2}\right)e^{i\eta_+}& 0\\
0 & 0 & 0 &1
\end{array}\right].
\end{equation}
The type of these nontrivial double-qubit gates is uniquely determined by the combination of parameters $\theta$, $\psi$ and $\phi_1-\phi_2$. For example, when $\psi=0$, $\phi_1=\phi_2$, $\theta=\pi/2$, we can have a SWAP-like gate up to an extra phase for the base $|ee\rangle$,
\begin{equation} \label{Uswap}
U\simeq\left[\begin{array}{cccc}
1 & 0 & 0 & 0\\
0 & 0 & 1 & 0\\
0 & 1 & 0 & 0\\
0 & 0 & 0 & -1
\end{array}\right].
\end{equation}
In the current scheme for the two-qubit gate, the counterdiabatic driving terms could be formulated in almost the same way as Eq.~(\ref{Hcd}) for the one-qubit gate. Nevertheless now the ancillary Hamiltonian $H_{TQD}(t)$ is written in the bases of $\{|b\rangle, |gg\rangle\}$. Still the Rabi frequencies $\Omega_1(t)$ and $\Omega_2(t)$ become exactly $\Omega_1^{TQD}(t)$ and $\Omega_2^{TQD}(t)$ in Eq.~(\ref{omegaM}), respectively.

Employing the same definition as in Eq.~(\ref{fone}), we can measure the performance of the double-qubit gates with or without TQD method by calculating the final average fidelity with a certain running time $T$ and the average fidelity dynamics. In this section, the initial state of the system can be generally written as $|\Psi(0)\rangle=\cos\alpha_1|gg\rangle+\sin\alpha_1\cos\alpha_2e^{i\alpha_3}|ge\rangle+\sin\alpha_1\sin\alpha_2e^{i\alpha_4}|eg\rangle$ with $\{\alpha_1,\alpha_2,\alpha_3,\alpha_4\}\in[0,2\pi]$. Here the parameter $E(t)$ is again set as $1$.

\begin{figure}[htbp]
\centering
\includegraphics[width=0.5\textwidth]{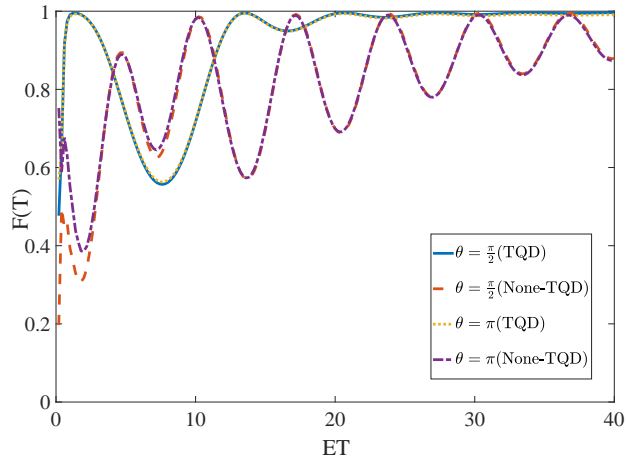}
\caption{(Color online) The effect of geometric transformation time $T$ on the final average fidelity $F(T)$ of the double-qubit gates under different parameter $\theta$. For $\theta=\pi/2$, the blue solid line and the red dashed line represent the fidelity obtained with and without the TQD approach, respectively. For $\theta=\pi$, the yellow dotted line and the purple dash-dotted line represent the fidelity obtained with and without the TQD, respectively. The other parameters are chosen as $\psi=0$, $\phi_1=\phi_2$, and $V=100E$.}\label{doublefidelity1}
\end{figure}

In Fig.~\ref{doublefidelity1}, we plot the final average fidelity of the double-qubit gates with or without the TQD under different $\theta$. As in Eq.~(\ref{omega}), $\theta$ is a parameter measuring the relative magnitudes of the two Rabi frequencies. Here the other parameters are fixed as $\psi=0$ (the phase difference between the two Rabi frequencies), $\phi_1=\phi_2$ (the control difficulty is relaxed since $\phi$ is invariant during the whole loop) and $V=100E$. It is found that the final average fidelity is roughly independent on the choice of $\theta$. The fluctuation amplitude of the two lines evaluated by the TQD approach (the blue and yellow lines) is apparently smaller than that of the two lines evaluated without the TQD approach (the red and purple lines). The first moment for the former two lines attaining unity is less than $ET=2$ while that for the latter two lines is greater than $ET=17$. The fidelity of the double-qubit gate is maintained almost unity by the CD terms over about $ET=20$. Comparing to the results for the single-qubit gate in Fig.~\ref{fidelity1}, the cyclic time required by the double-qubit gate is about ten times of that by the single-qubit gate. Without the help of the counterdiabatic term, it is hard to have a stable and high-fidelity transformation in a moderate running time from the red and purple lines, although both the fluctuation-amplitude and the quasi-period of the fidelity are gradually decreasing with $T$. Again it is therefore found that the TQD approach can dramatically reduce the running cost of the geometric gates.

\begin{figure}[htbp]
\centering
\includegraphics[width=0.5\textwidth]{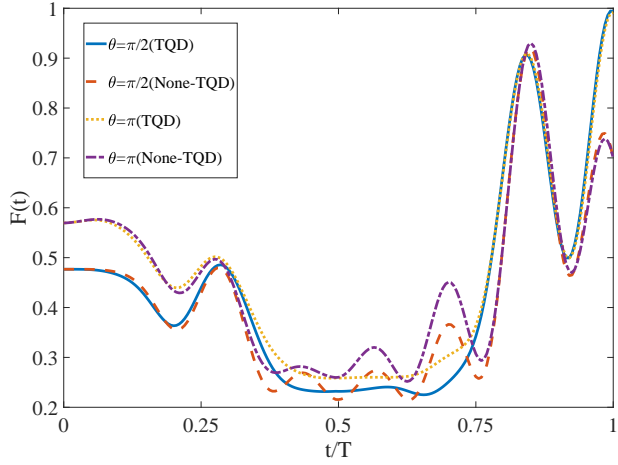}
\caption{(Color online) The dynamics of the average fidelity with a fixed cyclic time $ET=20$. For $\theta=\pi/2$, the blue solid line and the red dashed line represent the average fidelity dynamics with and without TQD approach, respectively. For $\theta=\pi$, the yellow dotted line the purple dash-dotted line represent the average fidelity dynamics with and without TQD, respectively. The other parameters are parameterized as $\psi=0$, $\phi_1=\phi_2$, $V=100E$.}\label{doublefidelity2}
\end{figure}

The fidelity dynamics of the quantum systems can also reflect the effect of the TQD method. In Fig.~\ref{doublefidelity2}, we plot the average-fidelity dynamics of the double-qubit gates under a fixed geometric transformation cyclic time $ET=20$ for $\theta=\pi/2$ and $\theta=\pi$. The effect upon the application of the TQD method (see the blue and yellow lines) is quite significant while the original Hamiltonian (see the red and purple lines) fails to maintain the adiabatic passage of the system in such a short cyclic time. The TQD approach enhances the final fidelity from about $0.8$ to nearly unity and greatly reduces the fluctuations during the dynamics. Therefore the TQD approach is indispensable to realize fast and high-fidelity double-qubit gates.

\begin{figure}[htbp]
\centering
\includegraphics[width=0.5\textwidth]{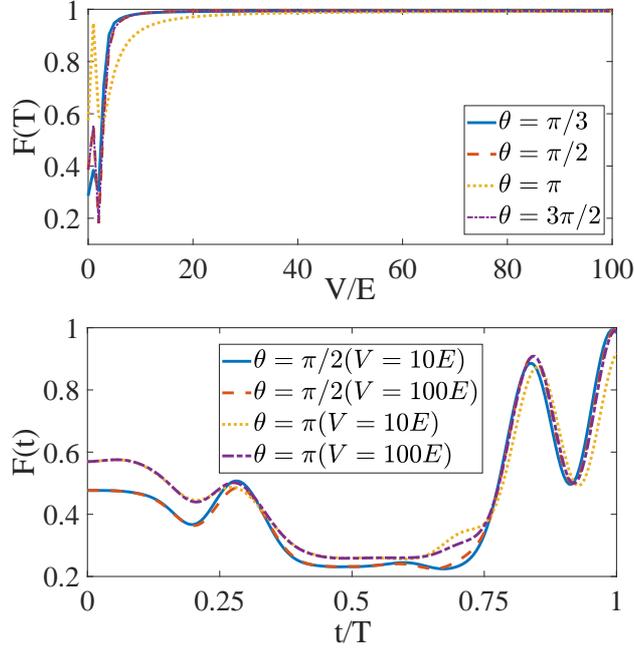}
\caption{(Color online) (a) The effect of the interaction strength $V$ on the final average fidelity $F(T)$ with different $\theta$. The blue solid line, the red dashed line, the yellow dotted line, and the purple dash-dotted line represent $\theta=\pi/3, \pi/2, \pi$ and $3\pi/2$, respectively. (b) The effect of the interaction on the dynamics of the average fidelity $F(t)$ with various pairs of $V$ and $\theta$. For $\theta=\pi/2$, the blue solid line and the red dotted line represent the average fidelity under $V=10E$ and $V=100E$, respectively. For $\theta=\pi$, yellow dotted line and purple dash-dotted line represent the average fidelity under $V=10E$ and $V=100E$, respectively. Other parameters are fixed as $\psi=0$, $\phi_1=\phi_2$, $ET=20$. }\label{V}
\end{figure}

The effect of coupling-strength $V$ is also measured by the final average fidelity and the average fidelity dynamics when we apply the TQD approach. In Fig.~\ref{V}(a), we plot the effect of $V$ on the final average fidelity under different $\theta$. We have a fixed cyclic time $ET=20$. It is found that the final average fidelity for $\theta=\pi/3$, $\theta=\pi/2$ and $\theta=3\pi/2$ will achieve $F(T)\approx1$ when $V\geq5E$. While under $\theta=\pi$, a nearly unity fidelity occurs when $V\geq30E$. A smaller $V$ gives rise to a much lower fidelity with fluctuations for all the cases. In Fig.~\ref{V}(b), We plot the average fidelity dynamics under $V=10E$ or $V=100E$ when the cyclic time is still set as $ET=20$. It is shown that for $\theta=\pi/2$, there is no apparent difference between a weak coupling strength with $V=10E$ and a strong one with $V=100E$. However, for $\theta=\pi$, the increment of $V$ from $10E$ to $100E$ yields the enhancement of the final fidelity from almost $0.9$ to $1$ at the final moment.

This scheme requires a strong interaction (about thirty times as large as the amplitude of driving fields) to realize a high-fidelity double-qubit gate. It can be realized in a practical scenario where two coupled Rydberg atoms enter the Rydberg-blockage regime. The Rydberg-mediated interaction between high-lying Rydberg states arising from dipole-dipole bond~\cite{dd1} or van der Waals forces can lead to the Rydberg blockade phenomenon~\cite{br1,br2,br3}, and it has been observed and demonstrated in many experiments~\cite{brn1,brn2}. When the Rydberg-mediated interaction is much stronger than the intensity of the driving pulses on the atoms, this blockade would prevent the simultaneous excitations of the neighboring Rydberg atoms~\cite{ra3}. It is so-called interaction-induced Rydberg blockade, meaning the simultaneous excitation of two atoms from their ground states to the Rydberg states is inhibited. This feature can be used to implement nontrivial double-qubit gates~\cite{ra3,tg1,tg2,tg3}. The Rydberg blockade can meet the requirement of our scheme. For example, in Ref.~\cite{dd1}, the magnitudes of $E$ is about $5\pi$MHz and the interaction can be taken up to $V\sim 200\pi$MHz. Also in the Rydberg blockade regime, the system is naturally robust against certain control errors~\cite{er1,er2}.

\begin{figure}[htbp]
\centering
\includegraphics[width=0.5\textwidth]{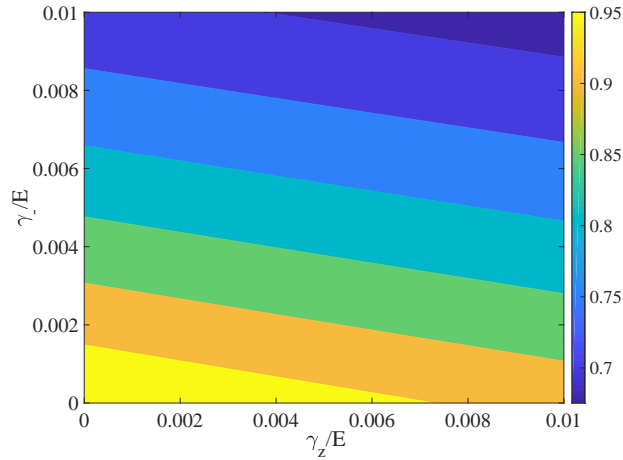}
\caption{(Color online) The effect of the system decoherence with decay rates $\gamma_-$ and $\gamma_z$ on the final fidelity of the two-qubit gate. Here the running time is fixed as $ET=20$, and the other parameters are chosen as $\psi=0$, $\phi_1=\phi_2$, $\theta=\pi/2$, and $V=100E$.}\label{doublegamma}
\end{figure}

Now we consider the effect of the system decoherence on the final fidelity after performing the TQD method. The decoherence including both energy dissipation and dephasing can be described by the following quantum master equation
\begin{eqnarray}
\label{ldtwo}
\frac{\partial\rho}{\partial t}=-i[H_{TQD}(t),\rho]+\frac{1}{2}\sum_{n\in\{1,2\}}\left[\Gamma_n^-L(\sigma^-_n)+\Gamma_n^zL(\sigma^z_n)\right].
\end{eqnarray}
Here $H_{TQD}(t)$ can be immediately obtained by inputting Eq.~(\ref{omegaM}) to the Hamiltonian~(\ref{Hmtwo}), and in the Lindbladian superoperator $L(A)$, $\sigma_n^-\equiv|g\rangle_n\langle e|$ and $\sigma_n^z\equiv|e\rangle_n\langle e|-|g\rangle_n\langle g|$, $n=1,2$. We assume that these two systems are in the common environments with the same decay rates, namely, $\Gamma^-_1=\Gamma^-_2=\gamma_-$ and $\Gamma^z_1=\Gamma^z_2=\gamma_z$. The effect of the dissipation by the rate $\gamma_-$ and the dephasing by the rate $\gamma_z$ of the system on the final fidelity of the two-qubit gate is plotted in Fig.~\ref{doublegamma}. In this case, the running time is set as $ET=20$ and the phase parameters are set as $\theta=\pi/2$, $\phi_1=\phi_2$, and $\psi=0$. At the initial moment, the atomic system is in the ground state, $|gg\rangle$. Then the fidelity $F(T)$ is defined as $F(T)=\langle gg|\rho(T)|gg\rangle$, where $\rho(T)$ represents the density operator $\rho$ at the final moment.

In Fig.~\ref{doublegamma} about the combined effect of both dissipation and dephasing on the double-qubit gate, it is found that the fidelity is insensitive to the dissipation noise indicated by $\gamma_-$ with a fixed dephasing noise indicated by $\gamma_z$. With the same magnitude of decay rates, the dissipation noise leads to about five times as the dephasing noise does with respect to the fidelity decline. The left-bottom triangle area indicates the parameter space for maintaining the double-qubit gate with high-fidelity $F(T)\geq0.95$. Obviously the double-qubit gate is more fragile to the external noise than the single-qubit gate (see Fig.~\ref{singlegamma2}).

\section{Nontrivial double-qubit gates II}\label{two2}

In this section, we propose an alternative scheme of nontrivial double-qubit gate by using a qubit (two-level system) to bridge two detuning-driven three-level systems.

\begin{figure}[htbp]
\centering
\includegraphics[width=0.3\textwidth]{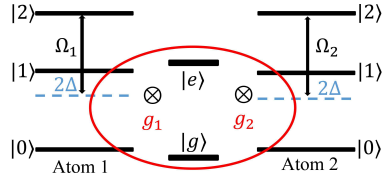}
\caption{(Color online) Diagram for two identical three-level systems coupled simultaneously to a two-level system (qubit). For either three-level system, the stable ground state $|0\rangle$ and the intermediate state $|1\rangle$ are coupled to the qubit with a coupling strength $g_n$, and the intermediate state $|1\rangle$ is coupled to the excited state $|2\rangle$ by an off-resonant driving fields with Rabi frequency $\Omega_n$, $n=1,2$. $2\Delta$ is the detuning between the driving frequency and the energy splitting between states $|1\rangle$ and $|2\rangle$.}\label{doublegate2}
\end{figure}

Consider two identical three-level atoms. Both of them are coupled with a two-level atom consisting of a ground state $|g\rangle$ and a high-energy state $|e\rangle$, as shown in Fig.~\ref{doublegate2}. For either three-level atom, the stable ground state $|0\rangle$ and the intermediate state $|1\rangle$ is coupled to the two-level system with the strength $g_n$, $n=1,2$. At the same time, for the $n$th three-level atom, the intermediate state $|1\rangle$ is coupled to the excited state $|2\rangle$ by an off-resonant and time-dependent driving $\Omega_n\exp[i\int^t_0ds\xi(s)]|2\rangle_n\langle1|+h.c.$, $n=1,2$, yielding a detuning $2\Delta\equiv\omega-\xi(t)$, where $\omega$ is the energy splitting between $|1\rangle$ and $|2\rangle$. Similar to Eq.~(\ref{Hone}), the total Hamiltonian of the quantum system in the rotating frame with respect to $U_0(t)=\exp[i\int_0^tds\xi(s)(|2\rangle_1\langle2|+|2\rangle_2\langle2|)]$ reads
\begin{eqnarray}\label{total}\nonumber
&H(t)=H_1+H_2(t), \\
&H_1=\left(g_1|0\rangle_1\langle1|+g_2|0\rangle_2\langle1|\right)\sigma^++h.c.,\\ \nonumber
&H_2(t)=\sum_{i=1,2}\big\{[\Omega_i(t)|2\rangle_i\langle1|+h.c.]-2\Delta(t)|1\rangle_i\langle1|\big\}
\end{eqnarray}
where $\sigma^+\equiv|e\rangle\langle g|$, $H_1$ represents the interaction Hamiltonian of the three-level atoms and the ancillary two-level atom, and $H_2(t)$ represents the Hamiltonian for the atom-laser interaction.

\begin{figure}[htbp]
\centering
\includegraphics[width=0.5\textwidth]{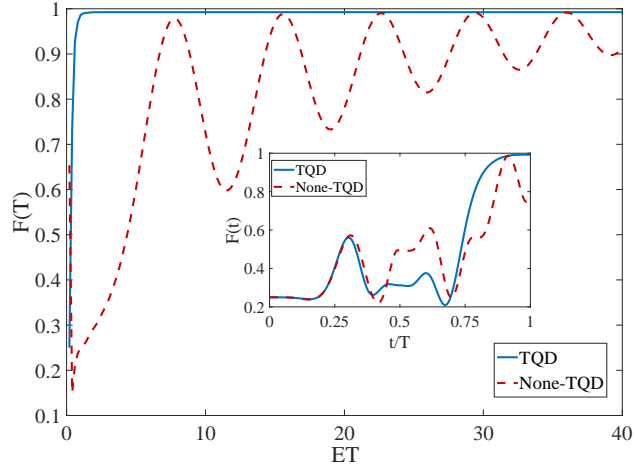}
\caption{(Color online) The final average fidelity and the fidelity dynamics (the inset with a fixed cyclic time $ET=20$) of the SWAP gate with (blue solid line) or without (red dashed line) the TQD. The other parameters are chosen as $\theta=3\pi/2$, $\phi_1=\phi_2$, and $g_1=g_2=100E$.}\label{Two}
\end{figure}

In the case that the intensities of the atom-atom coupling are significantly larger than that of the atom-laser coupling, namely, $g_{1,2}\gg\Omega_{1,2}(t)$ and the level spacing of the qubit system is far-off-resonant from all the other level spacings in the whole system, one can construct a double-qubit gate in a computational space spanned by $\{|00\rangle, |02\rangle, |20\rangle, |22\rangle\}$ as a subspace of the two three-level atoms. Strictly the computational space is spanned by $\{|00g\rangle, |02g\rangle, |20g\rangle, |22g\rangle\}$, where the qubit state $|g\rangle$ can be factored out. The details can be found in~\ref{App}. It is obvious that the evolution of the state $|00g\rangle$ is inhibited due to the fact that it is decoupled from the total system Hamiltonian~(\ref{total}). Also the evolution of the state $|22g\rangle$ is found to be irrelevant under the condition $g_{1,2}\gg\Omega_{1,2}(t)$. The effective Hamiltonian is found to be written in a subspace spanned by $\{|02g\rangle, |\varphi_0\rangle\equiv(g_2|10g\rangle-g_1|01g\rangle)/G, |20g\rangle\}$, where $G\equiv\sqrt{g_1^2+g_2^2}$, [see the analysis from Eq.~(\ref{fivespace}) to Eq.~(\ref{Htotaleff}) in~\ref{App}]
\begin{equation}\label{Hefftotal}
H_{eff}(t)=\left[\begin{array}{ccc}
0 & -\frac{\Omega_2(t)g_1}{G} & 0\\
-\frac{\Omega^*_2(t)g_1}{G} & -2\Delta(t) & \frac{\Omega^*_1(t)g_2}{G}\\
0 & \frac{\Omega_1(t)g_2}{G} & 0
\end{array}\right].
\end{equation}
It is not hard to see the effective Hamiltonian~(\ref{Hefftotal}) is in exactly the same form as the Hamiltonian~(\ref{Hone}) for the single-qubit gate. Thus similar to Eq.~(\ref{omega}), the elements can be parameterized as
\begin{eqnarray}\nonumber
-\Omega_2(t)g_1/G=\Omega(t)\sin(\theta/2)e^{-i\phi}, \\  \label{OG}
\Omega_1(t)g_2/G=\Omega(t)\cos(\theta/2)e^{-i\phi+i\psi},
\end{eqnarray}
with $\Omega(t)$ being real. Then in the frame spanned by $\{|b\rangle, |\varphi_0\rangle, |d\rangle\}$, where $|b\rangle$ and $|d\rangle$ are respectively the bright and dark states in this system and defined by mapping $|0\rangle\rightarrow|02g\rangle$ and $|2\rangle\rightarrow|20g\rangle$ in Eq.~(\ref{bd}), the Hamiltonian~(\ref{Hefftotal}) can be further expressed by
\begin{equation}\label{Hdouble}
H(t)=\Omega(t)\left(e^{-i\phi}|b\rangle\langle\varphi_0|+h.c.\right)-2\Delta(t)|\varphi_0\rangle\langle\varphi_0|.
\end{equation}
Repeating the procedure from Eq.~(\ref{Hone3}) to Eq.~(\ref{U}) while mapping $|1\rangle$ and $|b\rangle$ in Sec.~\ref{one} to $|\varphi_0\rangle$ and the new bright state $|b\rangle$, respectively, one can get a transformation operator in the same formation as in Eq.~(\ref{U}) yet in the new defined bases $\{|b\rangle,|d\rangle\}$. Plus the unchanged bases $|00g\rangle$ and $|22g\rangle$, one can obtain nontrivial double-qubit operators in the space spanned by $\{|00g\rangle,|d\rangle,|b\rangle,|22g\rangle$, i.e.,
\begin{equation}\label{Utwo}
\qquad U=\left[\begin{array}{cccc}
1 & 0 & 0 & 0\\
0 & 1 & 0 & 0\\
0 & 0 & e^{i\eta} & 0\\
0 & 0 & 0 & 1
\end{array}\right],
\end{equation}
where $\eta=\pi+\phi_1-\phi_2$, and $\phi_1$ and $\phi_2$ are respectively the parameters set in first-half and second-half stage of transformation period. Factoring out the common state $|g\rangle$ and then rotating to the subspace spanned by $\{|00\rangle, |02\rangle, |20\rangle, |22\rangle\}$, $U$ can be expressed by
\begin{equation}
\label{twoq2}
\qquad U=\left[\begin{array}{cccc}
1 & 0 & 0 & 0\\
0 & \cos^2\left(\frac{\theta}{2}\right)+\sin^2\left(\frac{\theta}{2}\right)e^{i\eta} & \frac{\sin\theta}{2}\left(e^{i\eta}-1\right)e^{-i\psi} & 0\\
0 & \frac{\sin\theta}{2}\left(e^{i\eta}-1\right)e^{i\psi} & \sin^2\left(\frac{\theta}{2}\right)+\cos^2\left(\frac{\theta}{2}\right)e^{i\eta}& 0\\
0 & 0 & 0 &1
\end{array}\right].
\end{equation}
A special combination of $\theta$, $\psi$, and $\eta$ then gives rise to a special double-qubit gate. For example, one can obtain a SWAP gate by letting $\theta=3\pi/2$, $\phi_1=\phi_2$, and $\psi=0$.

We can also apply the TQD method by adding the ancillary Hamiltonian $H_{TQD}(t)$ to the original Hamiltonian~(\ref{total}) to significantly reduce the running time for the desired adiabatic passages. It is straightforward to find that in the space $\{|02g\rangle, |\varphi_0\rangle, |20g\rangle\}$, the TQD term has the same form as in Eq.~(\ref{Hcd}). In particular, now the amplitudes and phases of two driving fields become
\begin{eqnarray}\nonumber
\Omega_1(t)&\to&\Omega^{TQD}_1(t)=-\frac{G}{g_1}[\Omega(t)-i\Lambda(t)]\cos(\theta/2)e^{-i\phi+i\psi}, \\ \label{omegamodify}
\Omega_2(t)&\to&\Omega^{TQD}_2(t)=\frac{G}{g_2}[\Omega(t)-i\Lambda(t)]\sin(\theta/2)e^{-i\phi}
\end{eqnarray}
where the parameters $\Lambda(t)$ and $E(t)$ share the same definitions as those in the single-qubit-gate case in Sec.~\ref{one}.

We take the SWAP gate as an example to measure the effect from the TQD method and demonstrate the performance of double-qubit gates. The initial state is assumed to be $|\Psi(0)\rangle=\cos\alpha_1|20g\rangle+\sin\alpha_1e^{i\alpha_2}|02g\rangle$ with $\{\alpha_1, \alpha_2\}\in[0,2\pi]$. The average fidelity we are interested is defined in Eq.~(\ref{fone}). The parameter $E(t)$ is set as $1$, and $\varphi(t)$ is also chosen as in Eq.~(\ref{varphi}).

In Fig.~\ref{Two}, the effect from the TQD approach can be observed by either the final fidelity under different running time $T$ or the fidelity dynamics during the geometric path with a fixed $T$. It is shown that the TQD approach render the final fidelity quickly attaining unity in less than $ET=1$. In contrast, without the TQD approach, the fidelity experiences significant fluctuations and pseudo-periodically attains unity although the amplitude of the fluctuation shrinks with increasing $T$. And the first time that the fidelity becomes nearly unity is about $ET=16$. In the inset of Fig.~\ref{Two}, we examine the dynamics of the double-qubit gates under a fixed geometric transformation period $ET=20$. The TQD approach enhances the final fidelity from about $0.8$ to nearly unity and significantly reduces the fluctuation amplitude during the time evolution. Therefore the TQD approach still plays an important role in obtaining stable and high-fidelity double-qubit gates.

\section{Discussion and Conclusion}\label{conc}

In summary, we have proposed a general scheme to realize a set of universal quantum superadiabatic geometric gates via the off-resonant-driven two or three level atoms. That can be impressively improved by the transitionless driving approach, in terms of the running time as well as the fidelity of the gates. The computational spaces of these gates, including a universal set of single-qubit gates and two schemes of nontrivial double-qubit gates, are encoded in the stable ground state and the long-lived excited states. In particular, for the single-qubit gates, we design a double-piecewise (super)adiabatic passages by two detuning driving fields and their phase difference. The two instantaneous eigenstates are individually deployed in either one of the two transformation-stages to cancel the accumulated dynamical phase when completing the whole loop. In the first scheme for the nontrivial double-qubit gate that can work in the strong coupling scenario, such as in the traditional Rydberg blockage regime~\cite{tg2,br3}, we employ a similar double-piecewise (super)adiabatic passages yet of both instantaneous eigenstates simultaneously to avoid the parallel transport condition. In the second scheme for the double-qubit gate, we employ a two-level system to mediate two three-level atoms under detuning driving.

Comparing to the previous works~\cite{nag,stag1,stag2,ra3}, our scheme has four distinguished features: (i) The time-dependent detuning between the driving frequencies and the atomic level-spacing is introduced as a controllable variable to avoid the requirement of parallel transport with respect to the vanishing dynamical phase, as well as the susceptibility to the systematic errors. (ii) The phase difference between the Rabi frequencies is used to realize an arbitrary single-qubit gate through merely a single-loop of the parametric space, which reduces the exposure time of the quantum device to the error sources. (iii) In the first scheme for nontrivial double-qubit gates, the two instantaneous eigenstates started individually from the ground state $|gg\rangle$ and the bright state $|b\rangle$ are simultaneously evolving under the effective Hamiltonian, which provides a novel geometric adiabatic procedure. (iv) In the second scheme for the double-qubit gates, a qubit system replaces the conventionally applied light-field to mediate the two detuning-driven three-level atoms, which enriches the implementation of double-qubit gates.

Our scheme is built up on the features of superadiabatic geometric gates and provides a promising approach to the high-fidelity quantum geometric computation.

\section*{Acknowledgments}

We acknowledge grant support from the National Science Foundation of China (Grants No. 11575071 and No. U1801661), Zhejiang Provincial Natural Science Foundation of China under Grant No. LD18A040001, and the Fundamental Research Funds for the Central Universities.

\appendix

\section{The effective Hamiltonian for the double-qubit gates in Sec.~\ref{two2}} \label{App}

This appendix is devoted to obtaining the effective Hamiltonian~(\ref{Hefftotal}) from the original Hamiltonian~(\ref{total}) in the main text. First, the amplitude of an arbitrary initial state on the basis $|00g\rangle$ will not change with time due to the fact that $H(t)|00g\rangle=0$ by Eq.~(\ref{total}). Second, if the initial state is populated at an arbitrary superposed state of $|02g\rangle$ and $|20g\rangle$, then the system will evolve in the subspace spanned by $\{|00e\rangle, |10g\rangle, |01g\rangle, |20g\rangle, |02g\rangle\}$. In this subspace, the total Hamiltonian~(\ref{total}) can be expressed by
\begin{equation}\label{fivespace}
H(t)=\left[\begin{array}{ccccc}
0 & g_1 & g_2 & 0 & 0\\
g_1 & -2\Delta(t) & 0 & \Omega_1^*(t) & 0\\
g_2 & 0 & -2\Delta(t) & 0 &\Omega_2^*(t)\\
0 & \Omega_1(t) & 0 & 0 & 0\\
0 & 0 & \Omega_2(t) & 0 & 0
\end{array}\right].
\end{equation}

The Hamiltonian~(\ref{fivespace}) can be simplified by using the eigenstates of $H_1$ in Eq.~(\ref{total}), which read [in the very subspace of Eq.~(\ref{fivespace})]
\begin{eqnarray}\label{eigenE}\nonumber
&|\varphi_0\rangle=\frac{1}{G}(g_2|10g\rangle-g_1|01g\rangle), \\
&|\varphi_{\pm}\rangle=\frac{1}{\sqrt{2}G}(g_1|10g\rangle+g_2|01g\rangle\pm G|00e\rangle).
\end{eqnarray}
The eigenvalues for $|\varphi_0\rangle$, $|\varphi_+\rangle$, and $|\varphi_-\rangle$ are $0$, $G$, and $-G$, respectively, where $G=\sqrt{g^2_1+g^2_2}$. Thus we have
\begin{equation}
\label{eigenH1}
H_1=G(|\varphi_+\rangle\langle\varphi_+|-|\varphi_-\rangle\langle\varphi_-|),
\end{equation}
and
\begin{eqnarray}\label{state}\nonumber
&|10g\rangle=\frac{1}{G}\big(g_2|\varphi_0\rangle+\frac{g_1}{\sqrt{2}}|\varphi_+\rangle+\frac{g_1}{\sqrt{2}}|\varphi_-\rangle\big), \\
&|01g\rangle=\frac{1}{G}\big(-g_1|\varphi_0\rangle+\frac{g_2}{\sqrt{2}}|\varphi_+\rangle+\frac{g_2}{\sqrt{2}}|\varphi_-\rangle\big).
\end{eqnarray}
In the same subspace as Eq.~(\ref{fivespace}), the Hamiltonian $H_2(t)$ of Eq.~(\ref{total}) can be written as [from now on $\Omega_n\equiv\Omega_n(t)$ and $\Delta\equiv\Delta(t)$ for simplicity]
\begin{eqnarray}
\label{subH2}
H_2(t)&=&(\Omega_1|20g\rangle\langle10g|+\Omega_2|02g\rangle\langle01g|+h.c.)-2\Delta|10g\rangle\langle10g|-2\Delta|01g\rangle\langle01g|.
\end{eqnarray}
Using Eq.~(\ref{eigenE}), $H_2(t)$ is rewritten as
\begin{eqnarray}
\label{eigenH2s}
\nonumber
H_2(t)&=&\frac{1}{G}\bigg(\Omega_1g_2|20g\rangle\langle \varphi_0|-\Omega_2g_1|02g\rangle\langle\varphi_0|+\frac{\Omega_1g_1}{\sqrt{2}}|20g\rangle\langle \varphi_+|\\ \nonumber
&+&\frac{\Omega_1g_1}{\sqrt{2}}|20g\rangle\langle\varphi_+|+ \frac{\Omega_2g_2}{\sqrt{2}}|02g\rangle\langle \varphi_-|+\frac{\Omega_2g_2}{\sqrt{2}}|02g\rangle\langle\varphi_-|+h.c.\bigg)\\ \nonumber
&-&2\Delta|\varphi_0\rangle\langle\varphi_0|-\Delta|\varphi_+\rangle\langle\varphi_+|-\Delta|\varphi_-\rangle\langle\varphi_+|-\Delta|\varphi_+\rangle\langle\varphi_-|-\Delta|\varphi_-\rangle\langle\varphi_-|.
\end{eqnarray}
Then in the rotating frame with respect to $U_0=e^{-iH_1t}$, the Hamiltonian~(\ref{fivespace}) becomes
\begin{eqnarray}
\label{eigenHtotal}\nonumber
H(t)&=&U_0(t)H(t)U_0^{\dag}(t)-iU_0(t)\dot{U}_0^{\dag}(t)\\ \nonumber
&=&\bigg(\frac{\Omega_1g_2}{G}|20g\rangle\langle \varphi_0|-\frac{\Omega_2g_1}{G}|02g\rangle\langle \varphi_0|+\frac{\Omega_1g_1}{\sqrt{2}G}|20g\rangle\langle\varphi_+|e^{iGt}\\\nonumber
&+&\frac{\Omega_1g_1}{\sqrt{2}G}|20g\rangle\langle \varphi_+|e^{iGt}+\frac{\Omega_2g_2}{\sqrt{2}G}|02g\rangle\langle \varphi_-|e^{-iGt}+\frac{\Omega_2g_2}{\sqrt{2}G}|02g\rangle\langle \varphi_-|e^{-iGt}+h.c.\bigg)\\ \nonumber
&-&2\Delta|\varphi_0\rangle\langle \varphi_0|-\Delta|\varphi_+\rangle\langle\varphi_+|-\Delta|\varphi_-\rangle\langle\varphi_-|-\Delta|\varphi_-\rangle\langle\varphi_+|e^{2iGt}-\Delta|\varphi_+\rangle\langle\varphi_-|e^{-2iGt}.
\end{eqnarray}
When $g_{1,2}\gg\Omega_{1,2}$, namely, $G\gg\Omega_{1,2}$, the terms involving $e^{\pm iGt}$ or $e^{\pm2iGt}$ can be omitted by the rotating wave approximation. So that in the subspace $\{|02g\rangle, |\varphi_0\rangle, |20g\rangle\}$, the effective Hamiltonian can be written as
\begin{eqnarray}
\label{Htotaleff}\nonumber
H_{eff}(t)&=&\frac{1}{G}\left(\Omega_1g_2|20g\rangle\langle\varphi_0|-\Omega_2g_1|02g\rangle\langle \varphi_0|+h.c.\right)-2\Delta|\varphi_0\rangle\langle\varphi_0|,
\end{eqnarray}
which is Eq.~(\ref{Hefftotal}) in the main text. One can check that both $|00g\rangle$ and $|22g\rangle$ are dark states of Eq.~(\ref{Htotaleff}) in nature and $|\varphi_0\rangle$ serves as an ancillary state to the computational space $\{|00\rangle, |02\rangle, |20\rangle, |22\rangle\}$.

However, we have to check the situation when the initial state is populated at $|22g\rangle$ since in the whole Hilbert space $H(t)|22g\rangle\neq0$ according to the full Hamiltonian~(\ref{total}). Actually in this case, the quantum system will evolve in the subspace spanned by $\{|01e\rangle,|02e\rangle,|10e\rangle,|11g\rangle,|12g\rangle,|20e\rangle,|21g\rangle,|22g\rangle\}$. In this subspace, the eigenstates and eigenvalues of the Hamiltonian $H_1$ are found to be
\begin{eqnarray}
\label{eigenEphi}\nonumber
&|\phi_0\rangle=\frac{1}{G}(g_1|10e\rangle-g_2|01e\rangle), \\
&|\phi_\pm\rangle=\frac{1}{\sqrt{2}G}(g_1|01e\rangle+g_2|10e\rangle\pm G|11g\rangle),
\end{eqnarray}
and $0$ and $\pm G$, respectively. Then we repeat the procedure from Eq.~(\ref{state}) to Eq.~(\ref{Htotaleff}). It turns out that the effective Hamiltonian in the subspace spanned by the new eigenstates of Eq.~(\ref{eigenEphi}) reads,
\begin{eqnarray}
\label{Htotaleff2}\nonumber
H_{eff}(t)&=&\frac{1}{G}\big(\Omega_1g_1|20e\rangle\langle\phi_0|-\Omega_2g_2|02e\rangle\langle\phi_0|+h.c.\big)-2\Delta|\phi_0\rangle\langle\phi_0|,
\end{eqnarray}
under the condition that $G\gg\Omega_{1,2}$. It is obvious that the state $|22g\rangle$ is decoupled from both effective Hamiltonians~(\ref{Htotaleff}) and (\ref{Htotaleff2}), so that it will remain invariant with the time evolution of the system. This result is consistent to that in the main text.

Note although the Hamiltonian~(\ref{Htotaleff2}) indicates intuitively that the subspace $\{|00e\rangle,|02e\rangle,|20e\rangle,|22e\rangle\}$ could be used as another computational subspace, yet the excited state $|e\rangle$ for the qubit is usually little populated. Thus normally we utilize the low-energy effective Hamiltonian~(\ref{Htotaleff}) in consideration of the stability of the geometric gates.

\end{document}